\documentstyle[art10]{article}
\begin{document}

\begin{titlepage}

\title{Octonion X-product and Octonion $E_{8}$ Lattices}

\author{Geoffrey Dixon\thanks{supported by the solidity of the planet earth,
 without
which none
of this work would have been possible.}
\\ Department of Mathematics or Physics \\ Brandeis University \\
Waltham, MA 02254 \\email:
dixon@binah.cc.brandeis.edu
\and Department of Mathematics \\ University of Massachusetts \\ Boston, MA
 02125}

\maketitle

\begin{abstract}
In this episode, it is shown how the octonion X-product is related to $E_{8}$
lattices, integral
domains, sphere fibrations, and other neat stuff.
\end{abstract}

\end{titlepage}

\section*{1. Introduction.}

Let {\bf O} be the octonion algebra {\bf [1]}, an 8-dimensional real division
algebra, both
noncommutative and nonassociative, and the last in the finite sequence of real
division algebras
including the reals, {\bf R}, complexes, {\bf C}, and quaternions, {\bf Q}.
Let $e_{a}, \;
a=0,1,...,7$, be a basis for {\bf O}, with
\begin{equation}
e_{0} = 1,
\end{equation}
the identity, and
\begin{equation}
(e_{a})^{2} = -1, \; a\in\{1,...,7\}.
\end{equation}
These latter elements also anticommute:
\begin{equation}
e_{a}e_{b} = -e_{b}e_{a}, \; a\ne b\in\{1,...,7\}.
\end{equation}
Finally, we choose an octonion multiplication whose quaternionic triples are
determined by the
cyclic product rule,
\begin{equation}
e_{a}e_{a+1} = e_{a+5}, \; a\in\{1,...,7\},
\end{equation}
where the indices in (4) are from 1 to 7, modulo 7 (and in particular we will
set 7=7 mod 7 to
avoid confusing $e_{0}$ with $e_{7}$).  Given (4), the following useful
property
is also satisfied:
\begin{equation}
e_{a}e_{b} = \pm e_{c} \; \; \Longrightarrow \; \; e_{2a}e_{2b} = \pm e_{2c},
\; \;
a,b,c\in\{1,...,7\}
\end{equation}
(this is the index doubling property; because of (4) and (5),  proofs of many
octonion properties
can be done very generally by proving the property in one example only).

Let $X\in {\bf O}$ be a unit element.  That is,
\begin{equation}
\parallel X \parallel^{2} = XX^{\dag} = (X^{0} + \sum_{a=1}^{7}X^{a}e_{a})
(X^{0} -
\sum_{a=1}^{7}X^{a}e_{a}) = \sum_{a=0}^{7}(X^{a})^{2} = 1.
\end{equation}
So,
\begin{equation}
X\in S^{7},
\end{equation}
the 7-sphere.  Because {\bf O} is nonassociative, if $X \in S^{7}$, and $A,B
\in
{\bf O}$ are
two other  elements, then
\begin{equation}
A\circ_{X}B \equiv (AX)(X^{\dagger}B) \ne AB
\end{equation}
in general.  However, if we fix $X$, then ${\bf O}_{X}$, which denotes {\bf O}
with its original
product replaced by the so-called X-product (8) (see {\bf [2]}, and also
{\bf [1][3]}), is yet
another copy of the octonions, isomorphic to the starting copy (or any other
copy).

In {\bf [3]} I showed that if $X \in \Xi_{0} \cup \Xi_{1} \cup\Xi_{2}
\cup\Xi_{3},$
where

\begin{equation}
\begin{array}{cl}
\Xi_{0} = & \{\pm e_{a}\}, \\ \\
\Xi_{1} = & \{(\pm e_{a}\pm e_{b})/\sqrt{2}: a,b \mbox{ distinct}\}, \\ \\
\Xi_{2} = & \{(\pm e_{a}\pm e_{b}\pm e_{c}\pm e_{d})/2: a,b,c,d
\mbox{ distinct}, \;
e_{a}(e_{b}(e_{c}e_{d}))=\pm 1\}, \\ \\
\Xi_{3} = & \{(\sum_{a=0}^{7}\pm e_{a})/\sqrt{8}:
\mbox{ odd number of +'s}  \}, \\ \\  & a,b,c,d\in\{0,...,7\}, \\
\end{array}
\end{equation} \\
then for all $a,b\in\{0,...,7\}$, there is some $c\in\{0,...,7\}$ such that
\begin{equation}
e_{a}\circ_{X}e_{b} = \pm e_{c}
\end{equation}
(in {\bf [3]} the superscript $+5$ was affixed to the sets $\Xi_{m}$ to
indicate
that the starting
multiplication was that determined by (4), but as I am using only this one
starting multiplication
here, I will dispense with the superscripts).  That is, in this case
${\bf O}_{X}$
can be obtained
from {\bf O} by a rearrangement of the indices in $\{1,...,7\}$ {\bf [3]}.

\section*{2. $E_{8}$ Lattices and Integral Domains.}

The 240 elements of $\Xi_{0} \cup \Xi_{2}$ are the nearest neighbors (first
shell)
 to the origin of an
$E_{8}$ lattice (so are the 240 elements of $\Xi_{1} \cup \Xi_{3}$ (see
{\bf [4][5]})).
 Define
\begin{equation}
{\cal E}_{8}^{h} = G^{h}[\Xi_{0} \cup \Xi_{2}], \; h=1,...,7,
\end{equation}
where $G^{h}$ is the $O(8)$ reflection taking $e_{0} \longleftrightarrow
e_{h}$.
These 7 sets are
nearest neighbor points for 7 different $E_{8}$ lattices, but in this case it
is well known
{\bf [4][5]} that the 240 points of ${\cal E}_{8}^{h}$, for each $h=1,...,7$,
close under
multiplication (however, because of nonassociativity they do not form a finite
group).  One may also
think of ${\cal E}_{8}^{h}$ as being the unital elements of a noncommutative
and nonassociative
integral domain.\\

It should be fairly obvious that if
$$
X\in {\cal E}_{8}^{h},
$$
then ${\cal E}_{8X}^{h}$, the X-product variant of ${\cal E}_{8}^{h}$, is also
closed under its
multiplication, since for all $A,B\in {\cal E}_{8}^{h}$,
$$
AX \in {\cal E}_{8}^{h} \; \mbox{ and } \; X^{\dag}B \in {\cal E}_{8}^{h}
\Longrightarrow
(AX)(X^{\dag}B) \in {\cal E}_{8}^{h}.
$$
(Clearly if $X\in {\cal E}_{8}^{h}$, then $X^{\dag} \in {\cal E}_{8}^{h}$.)
 However, only if $X$ is
also an element of $\Xi_{0} \cup \Xi_{2}$ will the resulting X-product also
satisfy (10).  For
example,
$$
{\cal E}_{8}^{7} \cap \Xi_{0} = \Xi_{0},
$$
and
$$
\begin{array}{cl}
{\cal E}_{8}^{7} \cap \Xi_{2} = &
\{(\pm 1 \pm e_{1} \pm e_{5} \pm e_{7})/2, \;
(\pm e_{2} \pm e_{3} \pm e_{4} \pm e_{6})/2, \\
 & (\pm 1 \pm e_{2} \pm e_{3} \pm e_{7})/2, \;
(\pm e_{4} \pm e_{6} \pm e_{1} \pm e_{5})/2, \\
 & (\pm 1 \pm e_{4} \pm e_{6} \pm e_{7})/2, \;
(\pm e_{1} \pm e_{5} \pm e_{1} \pm e_{5})/2\}. \\
\end{array}
$$
Since $\pm X$ results in the same X-product, there are 8 X-product variants
${\cal E}_{8X}^{7}$
arising from ${\cal E}_{8}^{7} \cap \Xi_{0}$, and $(6\times 16)/2 = 48$
arising from
${\cal E}_{8}^{7} \cap \Xi_{2}$.  So there are 56 X-product variants of
 ${\cal E}_{8}^{7}$ (and by
virtue of index cycling, all the ${\cal E}_{8}^{h}$) that satisfy (10),
and close under the new
X-product.  (There are clearly other X-product variants that close but do
 not satisfy (10), that is,
for which the resulting multiplication is not the result of a simple index
rearrangement {\bf [3]}.)

\section*{3. ${\cal E}_{8}^{0} \equiv \Xi_{0} \cup \Xi_{2}$ is Closed.}

Actually,
$$
{\cal E}_{8}^{0} \equiv \Xi_{0} \cup \Xi_{2}
$$
is not closed under the multiplication  we started with.  For example,
$$
(1 + e_{1} + e_{2} + e_{6})/2, \; (1 + e_{1} + e_{3} + e_{4})/2 \in
{\cal E}_{8}^{0},
$$
but
$$
(1 + e_{1} + e_{2} + e_{6})(1 + e_{1} + e_{3} + e_{4})/4 = (e_{1} + e_{2}
+ e_{4} + e_{5})/2
\not\in {\cal E}_{8}^{0},
$$
because
$$
e_{1}(e_{2}(e_{4}e_{5})) = e_{1}(e_{2}e_{2}) = -e_{1} \ne \pm 1.
$$

However, take a look at the ${\cal E}_{8}^{h}, h=1,...,7$.  These satisfy
\begin{equation}
{\cal E}_{8}^{h} = \Xi_{0} \cup \Xi_{2}^{h},
\end{equation}
where
\begin{equation}
\begin{array}{cl}
\Xi_{2}^{h} = &  \{(\pm e_{a}\pm e_{b}\pm e_{c}\pm e_{d})/2: a,b,c,d
\mbox{ distinct}, \\
 & e_{a}(e_{b}(e_{c}e_{d}))=\pm e_{h} \mbox{ if exactly one of } a,b,c,d= 0
\mbox{ or } h, \\
 & e_{a}(e_{b}(e_{c}e_{d}))=\pm 1 \mbox{ otherwise }\}.
\end{array}
\end{equation}

As it turns out, we can achieve very much the same thing on ${\cal E}_{8}^{0}$
using the X-product.
Consider
\begin{equation}
X = (1+e_{7})/\sqrt{2} \in \Xi_{1}.
\end{equation}
Let
\begin{equation}
A = (\pm e_{a}\pm e_{b}\pm e_{c}\pm e_{d})/2 \in \Xi_{2},
\end{equation}
so
\begin{equation}
e_{a}(e_{b}(e_{c}e_{d}))=\pm 1.
\end{equation}
If, however, we modify the product, using the X-product (8) with $X$ given
in (14), then the bits
of $A$ in (15) satisfying (16) also satisfy
\begin{equation}
\begin{array}{l}
e_{a}\circ_{X}(e_{b}\circ_{X}(e_{c}\circ_{X}e_{d})) = \pm e_{7} \\
\mbox{ if exactly one of } a,b,c,d= 0 \mbox{ or } h=7, \\
e_{a}(e_{b}(e_{c}e_{d}))=\pm 1
\mbox{ otherwise. } \\
\end{array}
\end{equation}
In other words (see (13)), ${\cal E}_{8}^{0}$ is closed under this particular
 X-product.  Note, for
example, that
\begin{equation}
(1 + e_{1} + e_{2} + e_{6})\circ_{X}(1 + e_{1} + e_{3} + e_{4})/4 = (e_{2}
 + e_{3} + e_{4} + e_{6})/2
\in \Xi_{2},
\end{equation}
since
$$
e_{2}(e_{3}(e_{4}e_{6})) = e_{2}(e_{3}e_{7}) = e_{2}(e_{2}) = -1
$$
(don't forget, $\Xi_{2}$ is defined in terms of the beginning product, not
the X-product variant).
Our principal result is then the following:
\begin{equation}
\fbox{$
{\cal E}_{8X}^{0} \mbox{ is X-product closed if } X \in \Xi_{1}.
$}
\end{equation}
Given that modulo sign change $\Xi_{1}$ has 56 elements, there are also
therefore 56
 X-products variants
${\cal E}_{8X}^{0}$ of ${\cal E}_{8}^{0}$ that are closed under multiplication,
 and from which we
may define integral domains.

\section*{4. Sphere Fibrations to Lattice Fibrations.}

Let
$$
X = X^{0}+X^{1}e_{1}+X^{2}e_{2}+X^{3}e_{3}+X^{4}e_{4}+X^{5}e_{5}+X^{6}e_{6}
+X^{7}e_{7}\in S^{7}.
$$
Then
\begin{equation}
\begin{array}{l} e_{1}\circ_{X} e_{2} = \\ \\  ((X^{0})^{2}+(X^{1})^{2}
+(X^{2})^{2}+(X^{6})^{2}
-(X^{3})^{2}-(X^{4})^{2}-(X^{5})^{2}-(X^{7})^{2})e_{6} \\ \\
  +2(X^{0}X^{5}+X^{1}X^{7}-X^{2}X^{4}+X^{3}X^{6})e_{3} \\ \\
  +2(-X^{0}X^{7}+X^{1}X^{5}+X^{2}X^{3}+X^{4}X^{6})e_{4} \\ \\
  +2(-X^{0}X^{3}-X^{1}X^{4}-X^{2}X^{7}+X^{5}X^{6})e_{5} \\ \\
  +2(X^{0}X^{4}-X^{1}X^{3}+X^{2}X^{5}+X^{7}X^{6})e_{7} \\ \\
= Y = Y^{6}e_{6} + Y^{3}e_{3} + Y^{4}e_{4} + Y^{5}e_{5} + Y^{7}e_{7} \\
\end{array}
\end{equation}
(see {\bf [1][3]}) defines a map from $S^{7} \longrightarrow S^{4}$.  That is,
\begin{equation}
(Y^{6})^{2} + (Y^{3})^{2} + (Y^{4})^{2} + (Y^{5})^{2} + (Y^{7})^{2} =1.
\end{equation}
Therefore, relative to this X-product, the set $\{e_{1}, \; e_{2}, \; Y\}$
is a quaternionic
triple.  This implies that the set of all
\begin{equation}
U = exp(\theta^{1}e_{1}+\theta^{2}e_{2}+\theta^{3}Y)
\end{equation}
is just $SU(2) \simeq S^{3}$, and
\begin{equation}
e_{1}\circ_{(UX)}e_{2} = (e_{1}\circ_{X}U)\circ_{X}(U^{\dag}\circ_{X}e_{2}) =
e_{1}\circ_{X}e_{2} = Y
\end{equation}
(see {\bf [3]}), since the $U$ and $U^{\dag}$ cancel each other out because
these two elements are
part of the quaternionic subalgebra of ${\bf O}_{X}$ generated by $e_{1}$ and
 $e_{2}$.  Therefore,
\begin{equation}
\{UX: \; U = exp(\theta^{1}e_{1}+\theta^{2}e_{2}+\theta^{3}Y)\} \simeq S^{3}
\end{equation}
is the $S^{3}$ fibre over $Y \in S^{4}$ in the exact sequence
\begin{equation}
S^{3} \longrightarrow S^{7} \longrightarrow S^{4},
\end{equation}
implicit in (20-24).  (Clearly the map (20) could be replaced by
$$
X \longrightarrow e_{a}\circ_{X} e_{b}, \; a\ne b \in \{1,...,7\};
$$
many other possibilities exist, which I will leave to the reader to explore.)

All of this translates to  lattices, the shells of which are  discrete versions
of  spheres.  In
particular, let
\begin{equation}
X \in {\cal E}_{8}^{0} \equiv \Xi_{0} \cup \Xi_{2} \subset S^{7}.
\end{equation}
Consider the map
\begin{equation}
X \longrightarrow e_{1}\circ_{X}e_{2}.
\end{equation}
Because of (10), the image of this map is the ten element set
\begin{equation}
{\cal Z}^{5} \equiv \{\pm e_{6}, \; \pm e_{3}, \; \pm e_{4}, \; \pm e_{5}, \;
\pm e_{7}\} \subset
S^{4},
\end{equation}
which is the inner shell of the 5-dimensional cubic lattice, ${\bf Z}^{5}$
(see {\bf [5]}).
Consider the fibre of elements of ${\cal E}_{8}^{0}$ mapping to $e_{6} \in
{\cal Z}^{5}$, which is
\begin{equation}
{\cal D}_{4} \equiv \{\pm 1, \; \pm e_{1}, \; \pm e_{2},  \;\pm e_{6}\}
\cup \{(\pm 1 \pm e_{1}\pm e_{2}\pm e_{6})/2\} \subset S^{3} \longrightarrow
 \{e_{6}\},
\end{equation}
which is the inner shell of a 24-dimensional $D_{4}$ lattice and integral
domain
(see {\bf [4][5]}).
Generalizing further from (25), we have an exact sequence
\begin{equation}
{\cal D}_{4}\subset S^{3} \longrightarrow {\cal E}_{8}^{0} \subset S^{7}
\longrightarrow {\cal Z}^{5} \subset S^{4}.
\end{equation}

\begin{itemize}
\item{NOTE: } $10\times 24 = 240$, which I leave to the reader to prove. \\
\end{itemize}

\section*{5. Conclusion.}

The motivation for this work is curiousity fired by the beauty of the
mathematics.
   I have
only scratched the surface of this vast interconnected mathematical realm, and
many of its
connections I will never see.  But I share a profound belief that the design of
our physical reality
is intimately linked to this web of mathematical notions {\bf [1]}, and this is
 my way of gaining a
better understanding of the web and its power.

\end{document}